\documentclass[12pt]{article}
\usepackage{epsfig}

\newcommand\pubnumber{UCLA/01/TEP/5}
\newcommand\pubdate{February 2000}
\newcommand\hepnumber{hep-th/0102186}

\def\Title#1{\begin{center} {\Large\bf #1 } \end{center}}
\def\Author#1{\begin{center}{ \sc #1} \end{center}}
\def\Address#1{\begin{center}{ \it #1} \end{center}}

\newcommand\pubblock{\rightline{\begin{tabular}{l} \pubnumber\\
         \pubdate\\ \hepnumber \end{tabular}}}
\newenvironment{Abstract}{\begin{quotation}  }{\end{quotation}}
\newenvironment{Presented}{\begin{quotation} \begin{center} 
             Presented at the\end{center}
      \begin{center}\begin{large}}{\end{large}\end{center} \end{quotation}}
\def\Acknowledgments{\bigskip  \bigskip \begin{center}
          \large\bf Acknowledgments\end{center}}

\makeatletter
\def\section{\@startsection{section}{0}{\z@}{5.5ex plus .5ex minus
 1.5ex}{2.3ex plus .2ex}{\large\bf}}
\def\subsection{\@startsection{subsection}{1}{\z@}{3.5ex plus .5ex minus
 1.5ex}{1.3ex plus .2ex}{\normalsize\bf}}
\def\subsubsection{\@startsection{subsubsection}{2}{\z@}{-3.5ex plus
-1ex minus  -.2ex}{2.3ex plus .2ex}{\normalsize\sl}}

\renewcommand{\@makecaption}[2]{%
   \vskip 10pt
   \setbox\@tempboxa\hbox{\small #1: #2}
   \ifdim \wd\@tempboxa >\hsize     
       \small #1: #2\par          
     \else                        
       \hbox to\hsize{\hfil\box\@tempboxa\hfil}
   \fi}

 \def\citenum#1{{\def\@cite##1##2{##1}\cite{#1}}}
 
\newcount\@tempcntc
\def\@citex[#1]#2{\if@filesw\immediate\write\@auxout{\string\citation{#2}}\fi
  \@tempcnta\z@\@tempcntb\m@ne\def\@citea{}\@cite{\@for\@citeb:=#2\do
    {\@ifundefined
       {b@\@citeb}{\@citeo\@tempcntb\m@ne\@citea\def\@citea{,}{\bf ?}\@warning
       {Citation `\@citeb' on page \thepage \space undefined}}%
    {\setbox\z@\hbox{\global\@tempcntc0\csname b@\@citeb\endcsname\relax}%
     \ifnum\@tempcntc=\z@ \@citeo\@tempcntb\m@ne
       \@citea\def\@citea{,}\hbox{\csname b@\@citeb\endcsname}%
     \else
      \advance\@tempcntb\@ne
      \ifnum\@tempcntb=\@tempcntc
      \else\advance\@tempcntb\m@ne\@citeo
      \@tempcnta\@tempcntc\@tempcntb\@tempcntc\fi\fi}}\@citeo}{#1}}
\def\@citeo{\ifnum\@tempcnta>\@tempcntb\else\@citea\def\@citea{,}%
  \ifnum\@tempcnta=\@tempcntb\the\@tempcnta\else
  {\advance\@tempcnta\@ne\ifnum\@tempcnta=\@tempcntb \else\def\@citea{--}\fi
    \advance\@tempcnta\m@ne\the\@tempcnta\@citea\the\@tempcntb}\fi\fi}
\makeatother

\def\I{{\cal I}}
\def\eps{\epsilon}
\def\e{\epsilon}
\def\Ord{{\cal O}}

\def\Tr{\mathop{\rm Tr}\nolimits}
\def\P{{\rm P}}
\def\NP{{\rm NP}}
\def\pol{\varepsilon}

\newbox\charbox
\newbox\slabox
\def\s#1{{      
        \setbox\charbox=\hbox{$#1$}
        \setbox\slabox=\hbox{$/$}
        \dimen\charbox=\ht\slabox
        \advance\dimen\charbox by -\dp\slabox
        \advance\dimen\charbox by -\ht\charbox
        \advance\dimen\charbox by \dp\charbox
        \divide\dimen\charbox by 2
        \raise-\dimen\charbox\hbox to \wd\charbox{\hss/\hss}
        \llap{$#1$}
}}

\def\spa#1.#2{\left\langle#1\,#2\right\rangle}
\def\spb#1.#2{\left[#1\,#2\right]}
\def\spab#1.#2.#3{\langle\mskip-1mu{#1}^- 
                  | #2 | {#3}^-\mskip-1mu\rangle}
\def\spba#1.#2.#3{\langle\mskip-1mu{#1}^+ 
                  | #2 | {#3}^+\mskip-1mu\rangle}
\def\spbb#1.#2.#3{\langle\mskip-1mu{#1}^+ 
                  | #2 | {#3}^-\mskip-1mu\rangle}
\def\lor#1.#2{\left(#1\,#2\right)}

\def\tree{{\rm tree}}
\def\oneloop{{1 \mbox{-} \rm loop}}
\def\twoloop{{2 \mbox{-} \rm loop}}

\def\eqn#1{eq.~(\ref{#1})}

\def\fig#1{fig.~{\ref{#1}}}

\newskip\humongous \humongous=0pt plus 1000pt minus 1000pt
\def\caja{\mathsurround=0pt}
\def\eqalign#1{\,\vcenter{\openup1\jot \caja
        \ialign{\strut \hfil$\displaystyle{##}$&$
        \displaystyle{{}##}$\hfil\crcr#1\crcr}}\,}
\newif\ifdtup

\newcounter{eqnumber}
\renewcommand{\theeqnumber}{\arabic{eqnumber}}
\def\equn{
\refstepcounter{eqnumber}
\eqno({\rm \theeqnumber})
}

\def\beq{\begin{equation}}
\def\eeq#1{\label{#1}\end{equation}}
\def\eeqn{\end{equation}}

\newenvironment{Eqnarray}%
   {\arraycolsep 0.14em\begin{eqnarray}}{\end{eqnarray}}
\def\beqa{\begin{Eqnarray}}
\def\eeqa#1{\label{#1}\end{Eqnarray}}
\def\eeqan{\end{Eqnarray}}

\let\bar=\overbar

\def\Dslash{\not{\hbox{\kern-4pt $D$}}}
\def\dslash{\not{\hbox{\kern-2pt $\del$}}}

\def\msb{{\bar{\ssstyle M \kern -1pt S}}}

\def\eps{\epsilon}

\def\s#1{\widetilde{#1}}

\def\lsim{\mathrel{\raise.3ex\hbox{$<$\kern-.75em\lower1ex\hbox{$\sim$}}}}
\def\gsim{\mathrel{\raise.3ex\hbox{$>$\kern-.75em\lower1ex\hbox{$\sim$}}}}

\begin{document}
\begin{titlepage}
\pubblock

\vfill
\def\thefootnote{\fnsymbol{footnote}}
\Title{Perturbative Quantization of Gravity Theories}
\vfill
\Author{Zvi Bern\footnote{Work supported by the
US Department of Energy  under grant DE-FG03-91ER40662.}} 

\Address{Department of Physics \\
         UCLA, Los Angeles, CA 90095 USA}
\vfill
\begin{Abstract}
We discuss relations between gravity and gauge theory tree amplitudes
that follow from string theory.  Together with $D$-dimensional
unitarity, these relations can be used to perturbatively quantize
gravity theories, i.e. they contain the necessary information for
calculating complete gravity $S$-matrices to any loop orders.  This
leads to a practical method for computing non-trivial gravity
$S$-matrix elements by relating them to much simpler gauge theory
ones. We also describe arguments that $N=8$ $D=4$ supergravity is less
divergent in the ultraviolet than previously thought.
\end{Abstract}
\vfill

\begin{Presented}
5th International Symposium on Radiative Corrections \\ 
(RADCOR--2000) \\[4pt]
Carmel CA, USA, 11--15 September, 2000
\end{Presented}
\vfill
\end{titlepage}
\def\thefootnote{\arabic{footnote}}
\setcounter{footnote}{0}
%


\section{Introduction}

In this talk we review 
work~\cite{BDDPR,AllPlusGrav,MHVGrav,Aaron,Square} that exploits
perturbative relations between gravity and gauge theories.  Although
both theories have a local symmetry, their dynamical behaviors
are quite different.  Nevertheless, in the context of
perturbation theory, it turns out that tree-level gravity amplitudes
can, roughly speaking, be expressed as a sum of products of gauge
theory amplitudes.  These tree-level relations between gravity and
gauge theory S-matrices are rather remarkable from a conventional
Lagrangian or Hamiltonian point of view but can be most
easily understood from the Kawai, Lewellen and Tye (KLT)~\cite{KLT}
relations between open and closed string tree amplitudes.  When
combined with the $D$-dimensional unitarity methods described in
refs.~\cite{SusyFour,Review}, it provides a new tool for investigating
the ultra-violet behavior of quantum gravity.  (The unitarity
methods have also been applied to QCD loop computations of
phenomenological interest and to supersymmetric gauge theory
computations~\cite{SusyFour,BRY,AllPlus2}.)

Ultraviolet properties are a central issue for perturbative gravity.
Although gravity is non-renormalizable by power counting, no
divergence has, in fact, been established by a direct calculation for
any four-dimensional supersymmetric theory of gravity.  Explicit
calculations have established that non-supersymmetric theories of
gravity with matter generically diverge at one
loop~\cite{tHooftVeltmanAnnPoin,tHooftGrav,DeserEtal}, and pure
gravity diverges at two loops~\cite{PureGravityInfinityGSV}.  However,
in any supergravity theory in $D=4$, supersymmetry Ward
identities~\cite{SWI} forbid all possible one-loop~\cite{OneLoopSUGRA}
and two-loop~\cite{Grisaru} counterterms.  Thus, at least a three-loop
calculation is required to directly address the question of
divergences in four-dimensional supergravity.  There is a candidate
counterterm at three loops for all supergravities including the
maximally extended version ($N=8$)~\cite{KalloshNeight,HoweStelle}.
However, no explicit three loop (super) gravity calculations have
appeared.  It is in principle possible that the coefficient of a
potential counterterm can vanish, especially if the full symmetry of
the theory is taken into account.  Based on explicit calculation, we
shall argue that this is indeed the case for the potential 
three-loop counterterm of $N=8$ supergravity.

With traditional perturbative approaches~\cite{DeWitt} to performing
explicit calculations, as the number of loops increases the number of
algebraic terms proliferates rapidly beyond the point where
computations are practical. We will take a different approach, relying
instead on a new formalism for perturbatively quantizing gravity.

\section{Method for Investigating Perturbative Gravity}

Our reformulation of quantum gravity is based on two ingredients:
\begin{enumerate}
\item  The Kawai, Lewellen and Tye relations between closed
and open string tree-level S-matrices~\cite{KLT}. 
\item  The observation that the $D$-dimensional tree amplitudes contain
all information necessary for building the complete perturbative
$S$-matrix to any loop order~\cite{SusyFour,Review}.
\end{enumerate}

\subsection{The KLT tree-level relations.}

In the field theory limit ($\alpha' \to 0$) the KLT relations for the four-
and five-point amplitudes are~\cite{KLT,BGK}
$$
\eqalign{
& M_4^{\rm tree}  (1,2,3,4) = 
 - i s_{12} A_4^{\rm tree} (1,2,3,4) \, A_4^{\rm tree}(1,2,4,3)\,, \cr
& M_5^{\rm tree}(1,2,3,4,5) = 
i s_{12} s_{34}  A_5^{\rm tree}(1,2,3,4,5)
                                     A_5^{\rm tree}(2,1,4,3,5) \cr
& \hskip 3 cm 
 + i s_{13}s_{24} A_5^{\rm tree}(1,3,2,4,5) \, A_5^{\rm tree}(3,1,4,2,5)\,,\cr}
\equn\label{KLTExamples}
$$
where the $M_n$'s are the amplitudes in a gravity theory stripped of
couplings, the $A_n$'s are the color-ordered gauge theory 
sub-amplitudes also stripped of couplings 
and $s_{ij}\equiv (k_i+k_j)^2$.  We suppress all $\pol_j$
polarizations and $k_j$ momenta, but keep the `$j$' labels to
distinguish the external legs. Full gauge theory amplitudes 
are given in terms of the partial amplitudes $A_n$, via
$$
\eqalign{
{\cal A}_n^{\rm tree}& (1,2,\ldots n) = 
 g^{(n-2)} \sum_{\sigma \in S_n/Z_n}
{\rm Tr}\left( T^{a_{\sigma(1)}} 
\cdots  T^{a_{\sigma(n)}} \right)
 A_n^{\rm tree}(\sigma(1), \ldots, \sigma(n)) \,, \cr} \hskip .6 cm 
$$
where $S_n/Z_n$ is the set of all permutations, but with cyclic
rotations removed, and $g$ is the gauge theory coupling constant.
The $T^{a_i}$ are fundamental representation
matrices for the Yang-Mills gauge group $SU(N_c)$, normalized so that
$\Tr(T^aT^b) = \delta^{ab}$.  
For states coupling with the strength of gravity, 
the full amplitudes including the gravitational coupling constant are, 
$$
{\cal M}_n^{\rm tree} (1,\ldots n) = 
\left({  \kappa \over 2} \right)^{(n-2)} 
M_n^{\rm tree}(1,\ldots n)\,, \hskip 1 cm 
$$
where $\kappa^2 = 32\pi G_N$.
The KLT equations generically hold for any closed string states, using
their Fock space factorization into pairs of open string states.

Berends, Giele and Kuijf~\cite{BGK} exploited the KLT
relations~(\ref{KLTExamples}) and their $n$-point generalizations to
obtain an infinite set of maximally helicity violating (MHV) graviton
tree amplitudes, using the known MHV Yang-Mills
amplitudes~\cite{ParkeTaylor}.  Cases of gauge theory coupled to
gravity have recently been discussed in ref.~\cite{Square}.
Interestingly, the color charges associated with any gauge fields
appearing in gravity theories are represented through the KLT
equations as flavor charges carried either by scalars or fermions.
For example, by applying the KLT equations the three-gluon
one-graviton amplitude may be expressed as
$$
\eqalign{
 {\cal M}_4^\tree  (1^-_g, 2^-_g, 3^+_g, 4^+_h)
 & =-i g {\kappa \over 2} 
s_{12} A_4^\tree(1^-_g, 2^-_g, 3^+_g, 4^+_g) \times
       A_4^\tree(1_s, 2_s, 4^+_g, 3_s) \hskip 2 cm \cr
& 
= g  {\kappa \over 2}  {\spa1.2^4 \over \spa1.2\spa2.3\spa3.4\spa4.1} 
\times \sqrt{2} f^{a_1 a_2a_3} \, {\spb4.3\spa3.2 \over \spa2.4} \,,\cr}
$$
where the $\pm$ superscripts denote the helicities and the subscripts
$h$, $g$ and $s$ denote whether a given leg is a graviton, gluon or
scalar.  On the right-hand side of the equation, the group theory
indices are flavor indices for the scalars.  On the left-hand side
they are reinterpreted as color indices for gluons. For simplicity,
the amplitudes have been expressed in terms of $D=4$ spinor inner
products (see e.g. ref.~\cite{ManganoReview}), although the
factorization of the amplitude into purely gauge theory amplitudes
holds in any dimension.  The spinor inner products are denoted by
$\spa{i}.j = \langle i^- | j^+\rangle$ and $\spb{i}.j = \langle i^+|
j^-\rangle$, where $|i^{\pm}\rangle$ are massless Weyl spinors of
momentum $k_i$, labeled with the sign of the helicity.  They are
antisymmetric, with norm $|\spa{i}.j| = |\spb{i}.j| = \sqrt{s_{ij}}$.


\subsection{Cut Construction of Loop Amplitudes}

We now outline the use of the KLT relations for computing multi-loop
gravity amplitudes, starting from gauge theory amplitudes. Although the KLT
equations hold only at the classical tree-level,
$D$-dimensional unitarity considerations can be used to extend them to
the quantum level.  The application of $D$-dimensional unitarity has
been extensively discussed for the case of gauge theory amplitudes
~\cite{SusyFour,Review}, so here we describe it only briefly.

\begin{figure}[b!]
\begin{center}
\epsfig{file=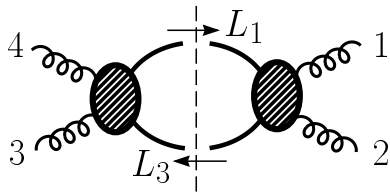,height=1.1in}
\caption[0]{The two-particle cut at one loop in the channel carrying 
momentum $k_1+k_2$.} 
\label{fig:TwoParticle}
\end{center}
\end{figure}

The unitarity cuts of a loop amplitude can be expressed in terms of
amplitudes containing fewer loops.  For example, the two-particle 
cut of a one-loop
four-point amplitude in the channel carrying momentum $k_1 + k_2$, as
shown in \fig{fig:TwoParticle}, can be expressed as the cut of,
$$
\eqalign{
 \sum_{\rm states} & \int {d^D L_1 \over (4 \pi)^D} \, 
  {i\over L_1^2} \,
{\cal M}_{4}^\tree(-L_1,1,2,L_3)
\,{i \over L_3^2}\,
{\cal M}_{4}^\tree (-L_3,3,4,L_1) \Bigr|_{\rm cut} \,, 
\cr}
\equn\label{BasicCutEquation}
$$
where $L_3 = L_1 - k_1 - k_2$, and the sum runs over all states
crossing the cut. We label $D$-dimensional momenta with capital
letters and four-dimensional ones with lower case.  We apply the
on-shell conditions $L_1^2 = L_3^2 = 0$ to the amplitudes appearing in
the cut even though the loop momentum is unrestricted; only functions
with a cut in the given channel under consideration are reliably
computed in this way.

Complete amplitudes are found by combining all cuts into a single
function with the correct cuts in all channels.  If one works with an
arbitrary dimension $D$ in \eqn{BasicCutEquation}, and takes care to
keep the full analytic behavior as a function of $D$, then the results
will be free of subtraction ambiguities that are commonly present in
cutting methods~\cite{ExactUnitarity,SusyFour,Review}. (The
regularization scheme dependence remains, of course.)  An advantage of
the cutting approach is that the gauge-invariant amplitudes on either
side of the cut may be simplified before attempting to evaluate the
cut integral~\cite{Review}.

\section{Recycling Gauge Theory Into Gravity Loop amplitudes}

As a relatively simple example, consider the one-loop amplitude with 
four identical helicity external gravitons and a scalar 
in the loop~\cite{AllPlusGrav,MHVGrav}. The cut in the $s_{12}$ channel is
$$
\eqalign{
\int {d^D L_1 \over (2\pi)^D} \; & {i \over L_1^2} 
 M_4^\tree(-L_1^s, 1_h^+, 2_h^+, L_3^s) \, {i \over L_3^2}
M_4^\tree(-L_3^s, 3_h^+, 4_h^+, L_1^s)\Bigr|_{\rm cut} \,, } \hskip 1 cm 
\equn\label{sChannelCut}
$$
where the superscript $s$ indicates that the cut lines are scalars
and the subscript $h$ indicates that the external particles are gravitons.
Using the KLT expressions (\ref{KLTExamples}) we may replace the
gravity tree amplitudes appearing in the cuts with products of gauge
theory amplitudes.  The required gauge theory tree amplitudes, with
two external scalar legs and two gluons, are relatively simple to
obtain using Feynman diagrams and are,
$$
\eqalign{
& A_4^\tree(-L_1^s,1_g^+,2_g^+,L_3^s) =  i
{\mu^2\spb1.2\over\spa1.2 [( \ell_1 -k_1)^2 -\mu^2] }\,, \cr
& A_4^\tree(-L_1^s, 1_g^+, L_3^s, 2_g^+) 
 = - i {\mu^2\spb1.2\over\spa1.2} 
 \biggl[{1\over (\ell_1 -k_1)^2 -\mu^2}
 + {1\over (\ell_1 -k_2)^2 -\mu^2}\biggr] \,, \cr}
$$
where $L_1 = \ell_1 + \mu$, where the subscript $g$ means the
lines are gluons.
The gluon momenta are four-dimensional, but the scalar momenta
are allowed to have a $(-2\e)$-dimensional component $\vec{\mu}$, with
$\vec{\mu}\cdot\vec{\mu} = \mu^2 > 0$.  The overall factor of $\mu^2$
appearing in these tree amplitudes means that they vanish in the
four-dimensional limit, in accord with a supersymmetry Ward
identity~\cite{SWI}.  In the KLT relation (\ref{KLTExamples}), one of
the propagators cancels, leaving
$$
\eqalign{
& M_4^\tree(-L_1^s, 1_h^+, 2_h^+, L_3^s) = 
- i \biggl({\mu^2\spb1.2\over\spa1.2}\biggr)^2 
 \biggl[{1\over (\ell_1 -k_1)^2-\mu^2}
 + {1\over (\ell_1 -k_2)^2-\mu^2}\biggr] \,. }
$$
By symmetry, the tree amplitudes appearing in any of the other cuts
are the same up to relabelings.  We then inserting these trees, with
appropriate leg labels, into the cut (\ref{sChannelCut}).

\begin{figure}[b!]
\begin{center}
\epsfig{file=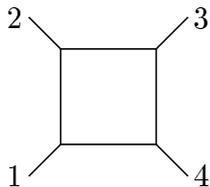,height=1in}
\caption[0]{The one loop box integral.}  
\label{OneLoopIntegralFigure}
\end{center}
\end{figure}

After combining all three cuts into a single function that has the correct
cuts in all channels one obtains the one-loop graviton amplitude with a
scalar in the loop,
$$
\eqalign{
M_4^\oneloop(1_h^+, 2_h^+, 3_h^+, 4_h^+)  &= 
    2 {\spb1.2^2\spb3.4^2\over\spa1.2^2\spa3.4^2} 
\Bigl(\I_4^\oneloop[\mu^8](s,t) + 
      \I_4^\oneloop[\mu^8](s,u) \cr
& \hskip 3 cm 
   + \I_4^\oneloop[\mu^8](t,u) \Bigr)\,, \cr } \hskip 2. cm 
\equn\label{FourGravAllPlus}
$$
where $s = s_{12}, \; t = s_{14}, \; u= s_{13}$ are the usual Mandelstam 
variables and
$$
\eqalign{
& \I_4^\oneloop[{\cal P}](s,t) = 
\int {d^D L \over (2\pi)^D} 
\, {{\cal P} \over L^2 (L - k_1)^2
       (L - k_1 - k_2)^2
        (L + k_4)^2  } \cr}
\equn\label{OneLoopIntegral} 
$$
is the scalar box integral depicted in \fig{OneLoopIntegralFigure}
with the external legs arranged in the order 1234. In
\eqn{FourGravAllPlus} the numerator $\cal P$ is $\mu^8$.  The two
other scalar integrals that appear correspond to the two other
distinct orderings of the four external legs.  The spinor factor
$\spb1.2^2\spb3.4^2/(\spa1.2^2\spa3.4^2)$ in \eqn{FourGravAllPlus} is
actually completely symmetric, although not manifestly so.  By
rewriting this factor and extracting the leading $\Ord(\eps^0)$
contribution from the integral, the final one-loop $D=4$ result after
reinserting the gravitational coupling is
$$
\eqalign{
{\cal M}_4^\oneloop &(1_h^+, 2_h^+, 3_h^+, 4_h^+)  = - \, {i \over (4 \pi)^2} 
\Bigl({\kappa\over 2} \Bigr)^4\,
\biggl( {st \over \spa1.2 \spa2.3 \spa3.4 \spa4.1} \biggr)^2 
{s^2 + t^2 + u^2 \over 120} \,, \cr}
\equn\label{FourGravPlus}
$$
in agreement with a previous calculation~\cite{DN}.

\section{Maximal Supergravity}

Maximal $N=8$ supergravity can be expected to be the least divergent of the
four-dimensional supergravity theories due to its high degree of
symmetry.  Moreover, from a technical viewpoint maximally
supersymmetric $N=8$ amplitudes are by far the easiest to deal with in
our formalism because of spectacular supersymmetric cancellations.
For these reasons it is logical to re-investigate the divergence
properties of this theory first~\cite{BDDPR}.  It
should be possible to apply similar methods to theories with less
supersymmetry.

\subsection{Cut Construction}

Again we obtain supergravity amplitudes by recycling gauge theory
calculations.  For the $N=8$ case, we factorize each of the 256 states
of the multiplet into a tensor product of $N=4$ super-Yang-Mills
states.  The key equation for obtaining the two-particle cuts is,
$$
\eqalign{
 \sum_{N=8 \atop \rm\ states}  
M_4^\tree(-L_1, & 1, 2, L_3) \times
  M_4^\tree(-L_3, 3, 4, L_1) \cr
&  =
s^2  \!\sum_{N=4 \atop \rm\ states} \!
  A_4^\tree(-L_1,  1, 2, L_3) \times
     A_4^\tree(-L_3,  3,4 , L_1)
     \cr
\null & \hskip .4 truecm
\times
\!\sum_{N=4 \atop \rm\ states} \!
 A_4^\tree(L_3, 1, 2 , -L_1) \times
                A_4^\tree(L_1, 3, 4, -L_3) \,,\cr}
\equn\label{GravitySewingStart}
$$
where we have suppressed the particle labels.  The external labels are
those for any particles in the supermultiplet, while the sum on the
left-hand side runs over all states in the $N=8$ super-multiplet.  On
the right-hand side the two sums run over the states of the $N=4$
super-Yang-Mills multiplet: a gluon, four Weyl fermions and six real
scalars.  Given the corresponding $N=4$ Yang-Mills two-particle sewing
equation~\cite{BRY},
$$
\eqalign{
 \sum_{N=4\atop \rm  states}
 A_4^\tree(-L_1, 1, & 2, L_3) \times
  A_4^\tree(-L_3, 3, 4, L_1)  \cr
&\hskip .3 cm 
 =  - i s t \, A_4^\tree(1, 2, 3, 4) \, 
   {1\over (L_1 - k_1)^2 } \, 
   {1\over (L_3 - k_3)^2 } \,, 
\cr} \hskip 2. cm 
$$
it is a simple matter to evaluate \eqn{GravitySewingStart}, yielding
$$
\eqalign{
 \sum_{N=8 \atop \rm\ states}  M_4^\tree(-L_1, 1, & 2, L_3) \times
  M_4^\tree(-L_3, 3, 4, L_1) \cr \
& \hskip .2 cm
= i stu M_4^\tree(1, 2, 3, 4)
 \biggl[{1\over (L_1 - k_1)^2 } + {1\over (L_1 - k_2)^2} \biggr] \cr
& \hskip 3.7 cm \times
\biggl[{1\over (L_3 - k_3)^2 } + {1\over (L_3 - k_4)^2} \biggr]\,. \cr}
\equn\label{BasicGravCutting}
$$
The sewing equations for the $t$ and $u$ channels are similar.

A remarkable feature of the cutting equation (\ref{BasicGravCutting})
is that the external-state dependence of the right-hand side is
entirely contained in the tree amplitude $M_4^{\rm \tree}$.  This fact
allows us to iterate the two-particle cut algebra to {\it all} loop
orders!  Although this is not sufficient to determine the
complete multi-loop four-point amplitudes, it does provide a wealth of
information.

Applying \eqn{BasicGravCutting} at one loop to each of the three
channels yields the one-loop four graviton amplitude of $N=8$
supergravity,
$$
\eqalign{
 {\cal M}_4^{\oneloop}(1, 2, 3, 4)
& =  -i \Bigl( {\kappa \over 2}\Bigr)^4 
s t u  M_4^\tree(1,2,3,4)  \cr
& \hskip .2 cm \times
 \Bigl(  \I_4^{\oneloop}(s,t) 
           + \I_4^{\oneloop}(s,u)  
           + \I_4^{\oneloop}(t,u)  \Bigr) \,, \cr} 
$$
in agreement with previous results~\cite{GSB}.  We have reinserted the
gravitational coupling $\kappa$ in this expression.  The scalar
integrals are defined in \eqn{OneLoopIntegral} with ${\cal P} = 1$.

\begin{figure}[b!]
\begin{center}
\epsfig{file=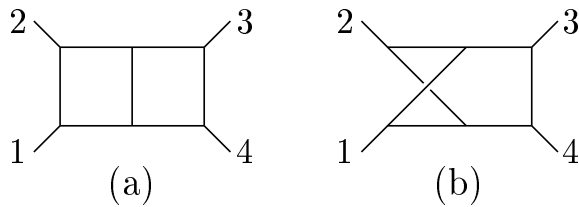,height=1.2in}
\caption[0]{The planar (a) and non-planar (b) scalar integrals, 
$\I_4^{\twoloop,\P}(s,t)$ and $ \I_4^{\twoloop,\NP}(s,t)$, appearing in
the two-loop $N=8$ amplitudes.  Each internal line
represents a scalar propagator.}  
\label{fig:PlanarNonPlanar}
\end{center}
\end{figure}

At two loops, the two-particle cuts are given by a simple iteration of
the one-loop calculation. The three-particle cuts can be obtained by
recycling the corresponding cuts for the case of $N=4$
super-Yang-Mills.  It turns out that the three-particle cuts introduce
no other functions than those already detected in the two-particle cuts.
Combining all the cuts into a single function yields the $N=8$
supergravity two-loop amplitude~\cite{BDDPR},
$$
\hskip -.15 cm 
\eqalign{ 
 {\cal M}_4^{\twoloop}(1,2,3,4) & = 
  \Bigl({\kappa \over 2} \Bigr)^6 \!\! stu  M_4^\tree(1,2,3,4) \cr
& \times
\Bigl(s^2 \, \I_4^{\twoloop,\P}(s,t) 
+ s^2 \, \I_4^{\twoloop,\P}(s,u)  \cr
& \hskip  .4 cm  
+ s^2 \, \I_4^{\twoloop,\NP}(s,t)
+ s^2 \, \I_4^{\twoloop,\NP}(s,u) 
+\;  \hbox{cyclic} \Bigr) \,, \cr}
\equn\label{TwoLoopAmplitude}
$$
where `$+$~cyclic' instructs one to add the two cyclic permutations of
legs (2,3,4), and $\I_4^{\twoloop,\P/\NP}$ are depicted
in \fig{fig:PlanarNonPlanar}.

We comment that using the two-loop amplitude (\ref{TwoLoopAmplitude}),
Green, Kwon and Vanhove~\cite{GreenTwoLoop} provided an explicit
demonstration of the non-trivial M theory duality between $D=11$
supergravity and type II string theory.  


\subsection{Divergence Properties of $N=8$ Supergravity}

Though a momentum cutoff scheme leads to a one-loop divergence for
$N=1$, $D=11$ supergravity, in dimensional regularization there are no
one-loop divergences in $D=11$, so the first potential divergence in
this theory is at two loops.  Dimensional regularization is a rather
convenient way to extract divergence properties as an analytic
function of dimension, allowing us to directly relate properties of
the $N=1$, $D=11$ supergravity to $N=8$ $D=4$ supergravity. (Some
care is needed, however, to preserve
supersymmetry~\cite{DimensionalReduction}.)

Since the two-loop $N=8$ supergravity amplitude
(\ref{TwoLoopAmplitude}) has been expressed in terms of scalar
$\phi^3$ loop momentum integrals, it is straightforward to extract the
divergence properties.  The scalar integrals diverge only for $D\ge
7$; hence the two-loop $N=8$ amplitude is manifestly finite in $D=5$
and $6$, contrary to earlier expectations based on superspace
power-counting arguments~\cite{HoweStelle}.  The discrepancy between
the above explicit results and the earlier superspace power counting
arguments is due to a previously unaccounted higher dimensional gauge
symmetry. Once this symmetry is accounted for, superspace power
counting gives the same degree of divergence as the explicit
calculation~\cite{StellePrivate}. 

The manifest $D$-independence of the cutting algebra allowed us to
extend the calculation to $D=11$, though there is no corresponding
$D=11$ super-Yang-Mills theory. The result (\ref{TwoLoopAmplitude})
then explicitly demonstrates that $N=1$ $D=11$ supergravity diverges
even when using dimensional regularization.  The $D=11$ two-loop
divergence may be extracted from the amplitude in
\eqn{TwoLoopAmplitude} yielding~\cite{BDDPR} a non-vanishing
counterterm.  Further work on the structure of the $D=11$ counterterm
has been carried out in refs.~\cite{DeserSeminara}.

Since the two-particle cut sewing equation iterates to all loop
orders, one can compute all contributions which can be assembled
solely from two-particle cuts~\cite{BDDPR}. Counting powers of loop
momenta in these contributions suggests the simple finiteness formula,
$$
L < {10\over (D-2)}\,,  \hskip 1 cm  \hbox{(with $L>1$),}
\equn\label{FinitenessBound}
$$
where $L$ is the number of loops.  This formula indicates that $N=8$
supergravity is finite in some other cases where the previous
superspace bounds suggest divergences~\cite{HoweStelle}, e.g. $D=4$,
$L=3$.  The first $D=4$ counterterm detected via the two-particle cuts
of four-point amplitudes occurs at five, not three loops.  Further
evidence that the finiteness formula is correct stems from the
maximally helicity violating contributions to $m$-particle cuts, in
which the same supersymmetry cancellations occur as for the
two-particle cuts~\cite{BDDPR}. Moreover, a recent superspace power
counting analysis taking the appropriate symmetries into account
confirms the finiteness bound~\cite{StellePrivate}.  Further work
would, however, be required to prove that there are no additional
hidden cancellations which could improve the finiteness condition
beyond \eqn{FinitenessBound}.  Interestingly, there has been a
suggestion by Chalmers that dualities might accomplish
this~\cite{ChalmersFinitenesss}. 

\section{Concluding Comments}

There are also a number of other interesting open questions. For example, the
methods described here have been used to investigate only maximal
supergravity. It would be interesting to systematically re-examine the
divergence structure of non-maximal theories.  (Some interesting recent work
on this may be found in ref.~\cite{DunbarJulia}.)  Using the 
methods described in this talk it might, for example, be possible to
systematically determine finiteness conditions order-by-order in the
loop expansion.  A direct derivation of the Kawai-Lewellen-Tye
decomposition of gravity amplitudes in terms of gauge theory ones
starting from the Einstein-Hilbert Lagrangian perhaps might lead to a
useful reformulation of gravity. Some initial steps to gain
an understanding of the Kawai-Lewellen-Tye relations, starting
from the Lagrangian was presented in ref.~\cite{Aaron}. (See also
ref.~\cite{Siegel}.)  Connected with this is the question
of whether the heuristic notion that gravity is the square of gauge
theory can be given meaning outside of perturbation theory.  In particular,
an intriguing question is whether it is possible to relate more
general solutions of the classical equations of motion for gravity to
those for gauge theory.


\Acknowledgments 
The research reported on here was supported by the
US Department of Energy under grants DE-FG03-91ER40662,
DE-AC03-76SF00515 and DE-FG02-97ER41029.  The work was
performed in collaborations with L. Dixon, D. Dunbar, A.K. Grant,
M. Perelstein, and J. Rozowsky.  An expanded version of this talk
appears in the proceedings of the Third Meeting on Constrained
Dynamics and Quantum Gravity, Villasimius (Sardinia, Italy) September
13-17, 1999.


\end{document}